# On Measuring Cognition and Cognitive Augmentation

Ron Fulbright

University of South Carolina Upstate
Spartanburg, SC USA
rfulbright@uscupstate.edu

**Abstract.** We are at the beginning of a new age in which artificial entities will perform significant amounts of high-level cognitive processing rivaling and even surpassing human thinking. The future belongs to those who can best collaborate with artificial cognitive entities achieving a high degree of cognitive augmentation. However, we currently lack theoretically grounded fundamental metrics able to describe human or artificial cognition much less augmented and combined cognition. How do we measure thinking, cognition, information, and knowledge in an implementation-independent way? How can we tell how much thinking an artificial entity does and how much is done by a human? How can we measure the combined and possible even emergent effect of humans working together with intelligent artificial entities? These are some of the challenges for researchers in this field. We first define a cognitive process as the transformation of data, information, knowledge, and wisdom. We then review several existing and emerging information metrics based on entropy, processing effort, quantum physics, emergent capacity, and human concept learning. We then discuss how these fail to answer the above questions and provide guidelines for future research.

## 1    Introduction

Until now, humans have had to do all of the thinking. However, we are at the beginning of a new era in human history in which artificial entities, we sometimes call "cogs" or "AIs," will perform greater amounts of high-level cognition rivaling or surpassing that of humans. The new era will see human cognitive performance augmented by working with such artificial entities. Although true machine intelligence has been predicted for decades, so far technology has fallen short of the grand vision. However, some impressive milestones have been achieved recently.

Chess was once considered the Holy Grail of artificial intelligence. However, in 1997, IBM's Deep Blue, defeated the reigning human chess champion [1]. Deep Blue was an expensive, specially-built computer system taking many years to develop. Now, even the simplest handheld electronic devices run Chess programs able to defeat all but the most advanced human players. Human chess players are achieving higher ratings than ever before by working with chess computers to refine their game. In this way, human chess players are already cognitively augmented. In 2011, a cognitive system built by IBM, called Watson, defeated the two most successful human champions of all time in the game of *Jeopardy!* [2][3][4]. Watson communicated in natural-language, demonstrated multi-level distributed reasoning, competed in real-time over being the first to ring the "buzzer," and engaged strategic wagering. In 2016, Google's AlphaGo defeated the reigning world champion in Go, a game vastly more complex than Chess [5][6]. In 2017, a version called AlphaGo Zero learned how to play Go by playing games with itself and not relying on any data from human games [7]. AlphaGo Zero exceeded the capabilities of AlphaGo in only three days. Also in 2017, a generalized version of the learning algorithm called AlphaZero was developed capable of learning any game. After only a few hours of self-training, AlphaZero achieved expert-level performance in the games of Chess, Go, and Shogi [8]. These herald a new type of artificial entity, one able to achieve, in a short amount of time, expert-level performance in a domain without special knowledge engineering or human input.

Of course, these systems are not built to just play games. Watson and AlphaGo represent a new kind of computer system built as a platform for a new kind of application [9][10]. For example, since 2011, IBM has been actively commercializing Watson technology to serve (and in many ways create) the emerging multi-billion dollar cognitive computing market. The Cognitive Business Solutions group consults with companies to create cogs. The Watson Health group's focus is to commercialize Watson technology for the health sector [11][12][13][14]. In her keynote address at the 2016 Consumer Electronics Show, Chairwoman, President, and CEO of IBM Ginni Rometty announced more than 500 partnerships with companies and organizations across 17 industries each building new applications and services utilizing cognitive computing technology based on Watson [16]. Many of these systems currently under development are intended for use by the average person.

In the coming age, many of us will encounter cognitive systems first via our handheld electronics. Also once deemed a pinnacle of artificial intelligence, natural language understanding and synthesis is now built into our voice-activated personal assistants such as Apple's Siri, Microsoft's Cortana, Google Now, Facebook's M, and Amazon Echo's Alexa [17][18][19][20][21]. The artificial entities we communicate with will rapidly increase in sophistication and cognitive ability. As cogs become able to perform higher-order cognitive processing, human-cog partnerships of the future will go far beyond what is possible today. Cogs will be able to consume vast quantities of unstructured data and information and deeply reason to arrive at novel conclusions and revelations, as well as, or better than, any human expert. Cogs will then become colleagues, co-workers, and confidants instead of tools. Because cogs will interact with us in natural language and be able to converse with us at human levels, humans will form relationships with cogs much like we do with friends, fellow workers, and family members. Fulbright has suggested such systems may very well lead to the democratization of expertise in much the same way the Internet has democratized information [22].

But what exactly is happening here? Obviously, these systems are processing information and generating information and knowledge that did not previously exist. However, do we have a way to measure how much knowledge has been created? Can we determine how much and by what quality the information has been altered? Do we have a way to compare one system's cognitive ability with another? How can we compare artificial cognition with human cognition? We predict humans will become cognitively augmented but how will we know when that happens? We do not yet have the metrics available to analyze cognition in this way. The purpose of this paper is to review some existing and emerging information and cognition metrics and suggest paths forward.

## 2 The Cognitive Process

The knowledge management and information science fields view data, information, knowledge, and wisdom (DIKW) as a hierarchy based on value as shown in Fig 1. [23]. Information is processed data, knowledge is processed information, and wisdom is processed knowledge. Each level is of a higher value than the level below it because of the processing involved.

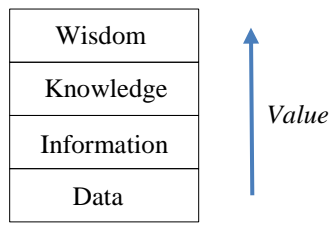

**Fig. 1.** The DIKW Hierarchy.

At the most fundamental level, a cognitive process transforms data, information, or knowledge, generically referred to as *information stock*, as depicted in Fig 2.

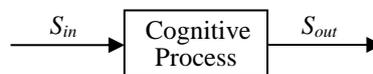

**Fig. 2.** A cognitive process as a transformation of information stock.

By viewing cognitive processing this way, the path to measuring cognition becomes a question of how to measure the effect a cognitive process has on the infor-mation stock and how to account the resources expended in performing the transformation. To measure this, we seek one or more characteristics altered by a cognitive process. Are there existing metrics we can use?

## 3   Previous Work

### 3.1   Entropy and Information Theory

There is a long history of entropic measures of information. In physics, particularly statistical mechanics and thermodynamics, such metrics are associated with the concept of order and disorder. In the mid-1800s, Clausius coined the term *entropy* to describe the amount of heat energy dissipated across the system boundary ultimately leading to the Second Law of Thermodynamics. In the late 1800s, Boltzmann related thermodynamic entropy of a system, S, to the number of equally likely arrangements (states) W, where $k$ is Boltzman's constant [24]

$$S = k \ln W. \tag{1}$$

Boltzmann's entropy is a measure of the amount of disorder in a system. In 1929, Leo Szilard was one of the first to examine the connection between thermodynamic entropy and information by analyzing the decrease in entropy in a thought experiment called Maxwell's Demon [25]. Szilard reasoned the reduction of entropy is compensated by a gain of information:

$$S = k \sum_i p_i \ln p_i, \tag{2}$$

where $p_i$ is the probability of the $i^{th}$ event, outcome, or state in the system. In 1944, Erwin Schrodinger wondered how biological systems (highly ordered systems) can become so structured, in apparent violation of the Second Law of Thermodynamics and realized the organism increases its order by decreasing the order of the environment [26]:

$$S = k \ln D \tag{3a}$$

Schrodinger calls -S, the negative entropy (or negentropy), a measure of order in a system. Leon Brillouin refined the idea and described living systems as importing and storing negentropy [27]. The ideas of Schrodinger, Szilard, and Brillouin involve a flow of information from one entity to another and use entropy to measure the flow.

The field of *information theory* was inspired by entropic measures. Ralph Hartley defined the information content, H, of a message of N symbols chosen from an alphabet of S symbols as [28]

$$H = \log S^N = N \log S \tag{4}$$

Since $S^N$ messages are possible, one can view the number of messages as the number of possible arrangements or states and therefore see the connection to thermodynamic entropy equations. Hartley's equation represents a measure of disorder in probability distribution across the possible messages, Hartley equates the measure of disorder with the information content of a message. In 1948, Claude Shannon developed the basis for what has become known as *information theory* [29][30][31]. Shannon's equation for entropy, H, is

$$H = -K \sum_{i=1}^{v} p(i) \log_2 p(i), \quad (5)$$

where p(i) is the probability of the i<sup>th</sup> symbol in a set of *v* symbols and K, is an arbitrary constant enabling the equation to yield any desired units. Shannon, as did Hartley, equates order/disorder and information content. Shannon entropy relies on the uncertainty of what the next symbol might be. The more unpredictable it is, the higher the Shannon information content for that symbol. The information content, *I*, of a message consisting of *m* symbols is

$$I = mH = -mK \sum_{i=1}^{m} p(i) \log_2 p(i) \quad (6)$$

Shannon's information theory has been used extensively for several decades and several other metrics have evolved from it including: *joint entropy* (measuring the uncertainty of a set of independent variables), *conditional entropy, mutual information* (measuring related information), *relative entropy* (measuring shared information), and a generalization of entropy metrics called *Renyi entropy*:

| | | |
|---|---|---|
| Joint Entropy | $H(X,Y) = -\sum_x \sum_y p(x,y) \log_2 p(x,y)$ | (7) |
| Conditional Entropy | $H(X\|Y) = -\sum_{x,y} p(x,y) \log_2 p(x\|y)$ | (8) |
| Mutual Information | $I(X;Y) = -\sum_{x,y} p(x,y) \log_2 \frac{p(x,y)}{p(x)p(y)}$ | (9) |
| Relative Entropy | $D_{KL}(p(X)\|\|q(X)) = \sum_x p(x) \log_2 \frac{p(x)}{q(x)}$ | (10) |
| Renyi Entropy | $H_\alpha(X) = \frac{1}{1-\alpha} \log_2 (\sum_{i=1}^{n} p_i^\alpha)$ | (11) |

where *p*(x,y) is the probability of the two occurring together, and *p*(x|y) is the probability of *x* given *y*. Different values of $\alpha$ in Renyi entropy yields other entropic metrics. $\alpha = 0$ yields Hartley entropy (Eq. 4). $\alpha = 1$ yields Shannon entropy (Eq. 5).

### 3.2 Algorithmic Information Content

In the 1960's, Ray Solomonoff, Gregory Chaitin, Andrey Kolomogorov and others developed the concept of *algorithmic information theory* (Kolmogorov-Chaitin complexity) as a measure of information [32][33][34][35]. The algorithmic information content, *I*, of a string of symbols, *w*, is defined as the size of the minimal program, s, running on the universal Turing machine generating the string

$$\mathbf{I}(w) = |\,\mathrm{s}\,|, \tag{12}$$

where the vertical bars indicate the length, or size, of the program *s*. This measure of information concerns the complexity of a data structure as measured by the amount of effort required to produce it. A string with regular patterns can be "compressed" and produced with fewer steps than a string of random symbols which requires a verbatim listing symbol by symbol. Like the entropic measures described above, this description equates order/disorder to information content, although in a different manner by focusing on the computational resources required.

### 3.3 Information Physics and Digital Physics

Starting with Szilard and Maxwell's Demon, described above, many have identified a deep connection between information and physical reality even describing information as having a physical manifestation in the universe. Shannon credited Szilard's work as his starting point in the 1950s. In 1967, Konrad Zuse suggested the universe itself is a computational structure, a notion now known as *digital physics.* Edward Fredkin was an early pioneer of digital physics maintaining all physical processes in nature are forms of computation or information processing. Rolf Landauer, for example, stated "information is physical" [36]. Stephen Wolfram has concluded the universe is digital in nature and can be described as simple programs [37]. Seth Lloyd has proposed the universe is a quantum computer and everything in it is "chunks of information." Lloyd maintains that merely by existing, physical systems register information and by evolving over time transform and process that information [38].

Digital physics permits us to state fundamental limits of information storage and computation. Since a bit of information requires a system to be in a particular state, the total number of bits a system can encode is limited by the total number of possible states. Furthermore, for information to be transformed, a system must move from one state to another. The Margolus/Levitin theorem implies the total number of elementary operations a system can perform per second is limited by its energy:

$$\#ops/sec \leq \frac{2E}{\pi \hbar} \tag{13}$$

where E is the system's average energy above the ground state and $\hbar$ is Planck's reduced constant [39]. The total number of bits available for a system to process is limited by its entropy:

$$\#bits \leq \frac{S}{k\ln 2} \tag{14}$$

where S is the system's thermodynamic entropy and *k* is Boltzmann's constant. The speed information can be moved from place to place is limited by the speed of light. Therefore, the maximum rate at which information can be moved in and out of a system with size R is

$$rate \approx \frac{cS}{kR} \tag{15}$$

attained by taking all the information S/kB ln2 in the system and moving it outward at the speed of light.

### 3.4 Emergent Capacity

While previous measures equated information content directly with entropy, in 1990, Tom Stonier suggested a more complex, exponential, relationship between entropy, *S*, and information, *I* [40][41][42]:

$$I = I_0 \, e^{-S/K} \tag{16}$$

where *K* is Boltzmann's constant, *S* is Shannon entropy, and $I_0$ is the amount of information in the system at zero entropy. Stonier maintained information content is in some way dependent on the *structure* present in a system and uses Shannon's entropy to provide the measure of that structure.

In 2002, Ron Fulbright explored the idea of information being an emergent property evolving from underlying complexity in a system [43]. Research in cellular automata and artificial life has shown emergence requires some randomness in the system. Systems that are too structured are not dynamic enough to allow structures to evolve. Systems that are too random are too dynamic to allow evolving structures to persist. Wolfram identified four classes of systems relating how dynamic evolving structures versus how random the system is [44]:

| | |
|---|---|
| **Class I:** | Patterns quickly degenerate into a homogeneous state. |
| **Class II:** | Simple static or periodic structures evolve. |
| **Class III:** | Increasingly random and chaotic patterns evolve. |
| **Class IV:** | Persistent complex structures evolve. |

Also studying cellular automata, Chris Langton demonstrated a phase change phenomenon by varying the randomness of cellular automata evolution and therefore identifying the ideal amount of randomness, $\lambda_c$, at which maximal emergent behavior evolves [45]. Langton discovered the ideal amount of randomness to be an intermediate value. Fig 3 shows the relationship between Langton's results and Wolfram's classes.

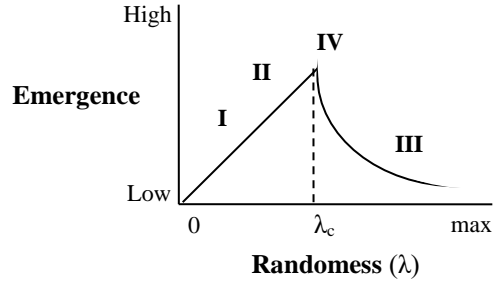

**Fig. 3.** Wolfram classes superimposed on Langton's phase change showing maximal emergent behavior occurring at an intermediate level of randomness.

Fulbright uses normalized entropy to measure a system's randomness and proposes the following equation for the capacity for emergent behavior by a system

$$I = me^{\eta log_2 \frac{1}{\eta}} \tag{17}$$

where $\eta = S/S_{max}$ and $m$ is the size of the system [43].

### 3.5 Representational Information Theory

Renaldo Vigo has recently proposed a new kind of information theory, *generalized representation information theory* (GRIT) [46][47]. Key to GRIT is how humans extract concepts (known as categories) from some given information (known as categorical stimuli). Elements in a categorical stimulus may vary from one another to various degrees and human concept extraction is based on the detection of patterns. Vigo's *generalized invariance structure theory* (GIST) maintains it is easier to extract a concept from information with less variance (more similarity between elements) than it is from information with a more variance (less similarity between elements). Values calculated with GIST formulae agree with empirical evidence from human trials lending further credence to the theory. GRIT defines concept-learning difficulty, called *structural complexity*, as

$$\psi(\pmb{F}) = |\pmb{F}|e^{-k\phi^2(\pmb{F})} \tag{18}$$

where $\pmb{F}$ is a well-defined category, $\phi$ is a measure of the categorical invariance of $\pmb{F}$, $|\pmb{F}|$ is the cardinality (size) of the category, and $k$ is a scaling parameter accounting for categories with different number of dimensions (e.g. $k = 2/D$ where D is the number of dimensions of category $\pmb{F}$). Subjective representational information is then the change in structural complexity due to some transformation as given by

$$h_s(F \rightarrow F') = \frac{\psi(F') - \psi(F)}{\psi(F)} = G \qquad (19)$$

where $F' \subseteq F$ and $F$ is transformed into $F'$. Later, we refer to this as *cognitive gain, G,* achieved by the transformation. Representational information, $h_s$, is defined as the percentage change in structural complexity effected by the transformation and ranges from -1 to +1. Positive values of $h_s$ indicate the learnability of the concept has gotten more difficult while negative values indicate the learnability of the concept has gotten easier. The representational information associated with any portion of the original can be calculated even down to the single element. Elements lowering structural complexity are more valuable than elements raising structural complexity.

Like Stonier and Fulbright's metric, this measure relates the value of information to the structure of the ensemble. However, Stonier and Fulbright's measures rely on entropic measures of patterns whereas in GRIT, patterns are measured by invariance.

### 3.6 Cognitive Augmentation Metrics

Building on GRIT, Fulbright has proposed several metrics useful for describing situations in which human cognition is augmented by the cognition of an artificial entity. Cognitive work, is an accounting of all changes in structural complexity caused by a transformation of some arbitrary information, $S$, keeping in mind some intermediate results may be deleted and therefore not represented in the output, $\psi_{lost}$ [48]. Cognitive work of a cognitive process is given by

$$W = |\psi(S_{out}) - \psi(S_{in})| + \psi_{lost} \qquad (20)$$

Cognitive work is a measure of the total effort expended in the execution of a cognitive process. It is important to note it requires both representational information (Eq. 19) and cognitive work to characterize a cognitive process.

When humans (H) and artificial entities, or cogs, (C) work together, each are responsible for some amount of change in representational information (cognitive gain (G)) and each expend a certain amount of cognitive work (W):

$$W_H = \sum_i W_H^i \qquad G_H = \sum_j G_H^j$$

$$W_C = \sum_i W_C^i \qquad G_C = \sum_j G_C^j$$

(21)

and the total amount by the ensemble is

$$W^* = W_H + W_C \qquad G^* = G_H + G_C \qquad (22)$$

Given that we can calculate the individual cognitive contributions, it is natural to compare their efforts. In fact, doing so yields the *augmentation factor, $A^+$*:

$$A_W^+ = \frac{W_C}{W_H} \qquad A_G^+ = \frac{G_C}{G_H} \qquad (23)$$

Note humans working alone without the aid of artificial entities are not augmented at all and have an $A^+ = 0$. If humans are performing more cognitive work than artificial entities, $A^+ < 1$. This is the world in which we have been living so far. However, when cogs start performing more cognitive work than humans, $A^+ > 1$ with no upward bound. That is the age that is coming.

Fulbright defines other efficiency metrics by comparing cognitive gain and cognitive work to each other and to other parameters such as time, $t$, and energy, $E$:

$$\text{Cognitive Efficiency:} \qquad \xi = \frac{G}{W} \qquad (24)$$

$$\text{Cognitive Power:} \qquad P_G = \frac{G}{t} \qquad P_W = \frac{W}{t} \qquad (25)$$

$$\text{Cognitive Density:} \qquad D_G = \frac{G}{E} \qquad D_W = \frac{W}{E} \qquad (26)$$

## 4 Discussion

Fig 2 represents cognition as the transformation of data, information, and knowledge (information stock) from a lower-value form to a higher-value form. If we can define a set of metrics to measure what the cognitive process does to the information stock and describe the effort/resources expended by the cognitive process, we could measure cognition to some degree. How well do the existing "information metrics" discussed above measure cognition and cognitive effort?

Information physics and digital physics metrics describe the universe at the quantum level. For the matter of measuring cognition and its effect on information stock, these metrics are too fine-grained. Imagine trying to assess the overall strategy of a sports team's performance by taking voltage measurements of individual neurons in the player's legs. We seek metrics of cognition working at the macro level where concepts, meaning, and semantics are the important, emergent, characteristics to measure.

Entropic-based metrics and algorithmic-based information content measures depend on the randomness (probability distribution) of the information stock. Does a cognitive process either increase or reduce the randomness of the information stock and if it does, does that increase the value of the information stock? Certainly, some cognitive processes seek to decrease randomness. We can imagine an alphabetization, or any other sorting algorithm, producing a much-less random output than its input. However, entropic measures and algorithmic measures attribute *less* information content to the ordered output than it does to the non-ordered input, so the metric operates opposite from what we expect. Furthermore, what about cognitive processes that do something other than change the order of the information? So, while entropic metrics measure a certain characteristic of information stock, they seem to fall short of a comprehensive measure of all cognitive processes.

The emergent-based metrics hold promise because they do not equate information content directly with randomness/order, rather they imply information content is the emergent result of structure having enough complexity to support a certain amount of information. However, Wolfram, Langton, and Fulbright studied cellular automata. The idea of information, knowledge, and wisdom being the emergent result of processing of lower-level information stock is appealing but the metrics need to be evolved so they calculate real and intuitive values for human-types of information stock. More research along these lines is encouraged.

Representational information theory is promising because it ties information content to human comprehension (learnability of a concept from categorical stimuli). If we seek to measure human-level cognition and processing of human-type of information stock, then a cognition metric should involve the human component. No other metric discussed here includes human understanding and meaning. However, it remains to be shown how rigorous statements about the *value* of information can be made using these metrics because they rely on *learnability* as the key quality and *invariance* as the key characteristic. Do these speak to value? Like the order/disorder discussion above for entropic metrics, some cognitive processes increase the learnability of a concept but other cognitive processes do not. Furthermore, some information is more important because of its implications and just measuring learnability is insufficient. For example, one can image two different categorical stimuli each with the same amount of invariance. Are these two of the same informational value? One could be samples of pizza toppings with the idea of identifying the best pizza while the other one may identify the type of cancer a patient has. Isn't the latter more valuable than the former? However, representational information metrics, nor any other metric, are not able to distinguish the two.

The current version of the cognitive augmentation metrics described above are based on representational information theory. However, they could be based on any other metric that measures the effect cognitive processing has on the information stock. One line of future research could seek to establish different fundamental metrics to underlie the cognitive augmentation metrics. Another line of future research could be to develop entirely new cognitive augmentation metrics not yet envisioned. Cognitive augmentation metrics have been shown to be able to differentiate between human and artificial cognition, they have not yet been used to solve an important problem in cognitive augmentation nor have they been used to predict results that could be tested in future research. We encourage researchers to both employ cognitive augmentation metrics in their research and seek to ground the theory in empirical reality.

A failure of all metrics discussed here is none speak to the *level* of cognition. Human cognition has been studied for decades and different levels of cognition have been defined. Bloom's Taxonomy is a famous hierarchy relating different kinds of cognition as shown below from easiest (remember) to hardest (create) [49].

- **Remember** (Recognize, Recall)
- **Understand** (Interpret, Exemplify, Classify/Categorize, Summarize, Infer/Deduce, Compare, Explain)
- **Apply** (Execute, Implement, Calculate)
- **Analyze** (Differentiate, Organize, Attribute)
- **Evaluate** (Check, Critique)
- **Create** (Generate, Plan, Produce)

**Fig. 4.** Bloom's Taxonomy expresses different levels of human cognitive processing

Each of these is a cognitive process effecting a transformation of information stock but the amount of effort involved increases dramatically as one goes down the list. Most cognitive systems and AIs today execute only the first few levels of Bloom's Taxonomy but cogs will quickly move into the higher-ordered types of processes. However, none of the metrics discussed here can distinguish between "easy" processing and "hard" processing. This is a critical line of inquiry for future research.

## 5 Conclusion

A new age is coming in which human cognition will be augmented by collaborating with artificial entities capable of high-level cognition. However, we do not yet have theoretically-grounded and empirically-grounded metrics to describe human or artificial cognition. For many years, the author thought a single metric could be developed to measure cognitive processes. However, current thinking is no single metric is possible and a family of metrics will have to be developed and verified empirically to fully characterize cognition. We have discussed several existing metrics from physics, information theory, and emergence theory. These metrics have deficiencies but each measure a certain characteristic of data, information, knowledge, and wisdom (information stock) such as order/disorder and learnability.

The family of metrics envisioned might very well employ the existing metrics discussed here but, future research must identify *all* important characteristics of information stock and devise ways to measure these characteristics. Researchers must not focus only on physical characteristics of the information stock. Things like value of information, importance, and emotional effect of information are human-oriented quantities possibly with subjective valuations. Researchers must include these and other human-centered issues in cognitive metrics. Finally, not all cognition is equal. We see this in human development. As a child ages, their cognitive abilities become more sophisticated, climbing the Bloom's Taxonomy hierarchy. As the coming cognitive systems age unfolds, artificial entities will master these higher-level cognitive processes. Researchers must devise metrics that consider level of cognition.